\documentclass[aps,prb,twocolumn]{revtex4}
\usepackage{graphicx}
\begin{document}

\title{Valence-bond crystal in a $\{111\}$ slice 
of the pyrochlore antiferromagnet}
\author{Oleg Tchernyshyov}
\author{Hong Yao}
\affiliation{Department of Physics and Astronomy, The Johns Hopkins University,
Baltimore, Maryland, 21218}
\author{R. Moessner}
\affiliation{Laboratoire de Physique Th\'eorique de l'Ecole Normale
Sup\'erieure, CNRS-UMR8549, Paris, France}

\begin{abstract}
We investigate theoretically the ordering effect of quantum spin
fluctuations in a Heisenberg antiferromagnet on a two-dimensional
network of corner sharing tetrahedra. This network is obtained as a
$\{111\}$ slice of the highly frustrated pyrochlore lattice, from
which it inherits the equivalence of all three pairs of opposite bonds
of each tetrahedron. The lowest-order (in $1/S$) quantum corrections
{\em partially} lift the huge degeneracy of the classical ground state and
select an ensemble of states with long-range valence-bond order.
\end{abstract}

\maketitle

Magnets with strong geometrical frustration have become a focus of
experimental studies and inspired a growing number of theoretical
inquiries into their unusual
properties.\cite{Schiffer96etal} For
classical spins, strong frustration precludes simple N\'eel order
and creates a vast ground state degeneracy. This degeneracy, in turn,
renders the magnet susceptible to nominally small perturbations
(dipolar interactions, anisotropies, quantum effects etc.).

An antiferromagnet on the pyrochlore lattice---a network of
corner-sharing tetrahedra---is perhaps the ultimate example of strong
frustration.\cite{oldpyro,Moessner98} Monte Carlo simulations with
classical Heisenberg spins show that thermal fluctuations alone are
ineffective in restoring magnetic order: the magnet remains in a
paramagnetic but correlated state at a temperature as low as
$10^{-4} JS^2$.  The effect of quantum fluctuations on the pyrochlore
antiferromagnet has not yet been
determined. 

Previous studies have focused on the case of maximal quantum
fluctuations, $S=1/2$. Early on, a ground state in the form of a
valence-bond crystal (VBC) was proposed,\cite{Harris91} which preserves
the spin-rotational $O(3)$ symmetry but breaks a number of discrete
lattice symmetries, including the equivalence of the three pairs of
opposite bonds of a tetrahedron. The relevant local order parameter is
constructed from the expectation values $\langle {\bf S}_i \cdot {\bf
S}_j\rangle$. This result was obtained, in the absence (to this day)
of a demonstrably reliable analytical treatment, via an uncontrolled
approximation; several methods have since yielded similar
results.\cite{Tsunetsugu01etal}

An alternative route of studying quantum effects begins in the
classical limit $S \to \infty$. At large $S$, quantum fluctuations are
weak and can be evaluated in the framework of a perturbation theory,
wherein physical quantities are expanded in powers of $1/S$.  The
large-$S$ approach has been previously applied to a variety of
strongly frustrated magnets.\cite{Chalker91,Yildirim96,Sobral97,Tch03}

\begin{figure}
\centerline{\includegraphics[width=\columnwidth]{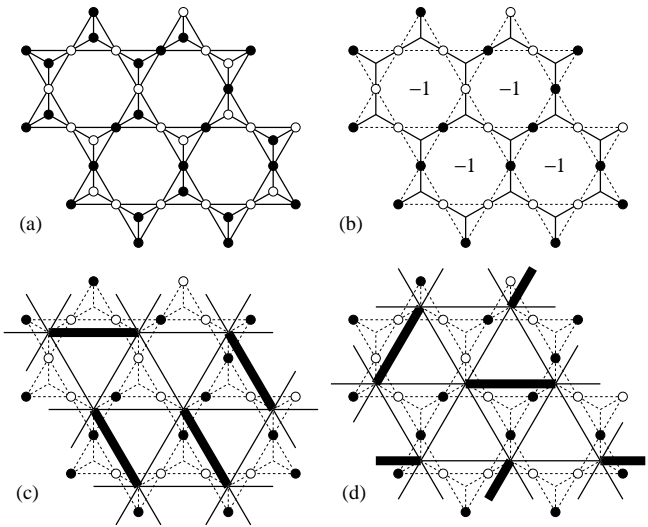}}
\caption{(a) A generic collinear ground state of Eq.~(\ref{eq-H}) on
the pyrochlore slice.  Open (filled) dots correspond to $s_i = +1$
($-1$).  (b) Kagome sites live on links of a honeycomb lattice.  The
$Z_2$ flux is $-1$ through every hexagon.  (c) and (d) Resulting dimer
coverings of the two triangular sublattices.  The dimers cross
frustrated bonds ($s_i = s_j$).}
\label{fig-wafer}
\end{figure}

In this paper, we report an application of the large-$S$ method to a
$\{111\}$ slice of the pyrochlore lattice shown in
Fig.~\ref{fig-wafer}(a).  The resulting lattice can be described as a
kagome plane flanked by two triangular lattices, so that every
triangle is promoted to a tetrahedron.  (Note that this is different
from the pyrochlore slab geometry of SCGO,\cite{SCGO} where a
triangular lattice is sandwiched between two kagome planes.)

From the symmetry point of view, our $\{111\}$ slice has the following
features important in connection with the pyrochlore in 3D. All
its units are equivalent tetrahedra and it thus permits collinear
classical ground states (unlike SCGO). Further, unlike the
checkerboard version of the 2D pyrochlore,\cite{Tch03} it
retains equivalence between all three pairs of opposing bonds of
each tetrahedron, a symmetry spontaneously broken in the $S=1/2$
pyrochlore VBC mentioned above.\cite{Harris91}

For this lattice, we have found and characterized the subset of classical
ground states favored by quantum fluctuations. The problem of minimizing
the zero-point energy is equivalent to a discrete gauge-like theory, which
by its very nature has many degenerate ground states. A natural first
guess is that the ground states are those with uniform flux; this ensemble
leads to a semiclassical valence bond liquid. However, our numerical
minimisations reveal vacua with a pattern of gauge fluxes violating
translational symmetry; this is reflected in a spatial modulation of
nearest-neighbor spin correlations $\langle {\bf S}_i \cdot {\bf S}_i
\rangle$.  Along with the absence of N\'eel order, this implies that the
spin system is a valence-bond crystal.

We study the Heisenberg spin Hamiltonian
\begin{equation}
H = J \sum_{\langle ij \rangle} {\bf S}_i \cdot {\bf S}_j
= (J/2)\sum_{\alpha} |{\bf L}_\alpha|^2 + {\rm const}
= \mathcal O(S^2),
\label{eq-H}
\end{equation}
where $J>0$ and the sum is taken over all bonds $(ij)$ shown in
Fig.~\ref{fig-wafer}(a).  In the alternative representation, the Greek
index $\alpha$ enumerates tetrahedra; ${\bf L}_\alpha$ is the sum of
the four spins residing on tetrahedron $\alpha$.  The energy
(\ref{eq-H}) is minimized when ${\bf L}_\alpha = 0$ for every
tetrahedron $\alpha$.  Classically, this condition does not determine
a unique ground state.  (This is the hallmark of frustration.)
Rather, it defines an enormous manifold of classical vacua. By
explicit construction, or Maxwellian counting,\cite{Moessner98} the
dimension of the manifold can be seen to equal twice the number of
tetrahedra. Thermal order by disorder is absent.

Quantum fluctuations produce virtual excitations out of the ground
state, thereby lowering the vacuum energy.  For large $S$, the quantum
correction can be expanded in powers of $1/S$ by expressing the spin
operators in terms of Holstein--Primakoff bosons.  (The virtual
excitations are magnon pairs produced by the operators $b_i^\dagger
b_j^\dagger$ in the bosonized Hamiltonian.)  The lowest-order
correction is the zero-point energy of harmonic spin waves,
\begin{equation}
E^{(1)} = -2 N J S + \sum_{a} \hbar |\omega_a|/2 = \mathcal O(S),
\label{eq-E1}
\end{equation}
where $\{\omega_a\}$ is the set of spin-wave frequencies in a given
classical ground state and $N$ is the number of tetrahedra.  The first
quantum correction (\ref{eq-E1}) depends implicitly (through the
magnon spectrum) on the spin configuration and thus can be viewed as
an effective spin potential breaking the degeneracy of the
ground-state manifold.

Typically, quantum fluctuations select collinear ground
states,\cite{Henley87} provided that such vacua exist at the classical
level.  Although we are unaware of any general proof of this
proposition, a compelling argument in its favor can be stated
nonetheless.  Given that the reduction of the vacuum energy is due to
generation of virtual magnon pairs and that magnons are transverse
excitations, the zero-point energy should be lower when more adjacent
spins share both of their transverse directions, i.e. when spins are
collinear.  In certain cases this heuristic argument can be made
precise.\cite{Yildirim96,Tch03,Chayes03} We assume that it applies in
this case as well and minimize the zero-point energy in the set of
collinear classical vacua.  As a quick check, 
we have computed zero-point energies
of a few simple classical vacua with the smallest magnetic unit cell,
and found them to be $-1.2126JS$, $-1.1612JS$, and $-1.1436JS$ for the
collinear, ``perpendicular,'' and ``anticollinear'' states,
respectively (notation of Ref.~\onlinecite{Henley87}).

We parametrize collinear states by means of Ising variables $s_i = \pm
1$ such that ${\bf S}_i/S = s_i \hat{\bf n}$, where $\hat{\bf n}$ is a
unit vector.  The constraint of zero total spin on every tetrahedron
carries over to the Ising variables: $s_0 + s_1 + s_2 + s_3 = 0$.  We
can eliminate the variable $s_0$ on the out-of-plane site at the
expense of introducing another constraint:
\begin{equation}
s_1 + s_2 + s_3 = \pm 1
\mbox{ on every triangle}.
\label{eq-constraint} 
\end{equation}
It follows from Eq.~(\ref{eq-constraint}) that collinear classical
vacua of the pyrochlore slice can be mapped onto the ground states of
a kagome Ising antiferromagnet.  Redundancy of the out-of-plane Ising
variables allows us to speak interchangeably of triangles and
tetrahedra when we deal with ground-state configurations.

The zero-point energy (\ref{eq-E1}) is an implicit function of the
Ising variables $\{s_i\}$. As outlined in Refs.~\onlinecite{Moessner98} 
and \onlinecite{Tch03}, the spin-wave spectrum can be
obtained by linearizing the classical equations of motion for the
spins, written in terms of the ${\bf L}$ variables: $\hbar \dot{\bf
L}_\alpha = \sum_{\beta} {\bf S}_{\alpha\beta} \times {\bf
L}_{\beta}.$ Here ${\bf S}_{\alpha\beta}$ denotes the spin shared by
tetrahedra $\alpha$ and $\beta$.  (All of these spins reside in the
kagome plane.)  For a collinear spin state polarized along $\hat{\bf
n} = (0,0,1)$, transverse fluctuations of the total spin on
tetrahedron $\alpha$ can be parametrized by a complex number
$\sigma_\alpha$ as ${\bf L}_\alpha = 
({\rm Re} \sigma_\alpha, {\rm Im}
\sigma_\alpha, 0)$.  The magnon frequencies are then determined by
solving the eigenvalue problem
\begin{equation}
J S \sum_{\beta} s_{\alpha\beta}\, \sigma_\beta 
= \hbar \omega \, \sigma_\alpha.
\label{eq-omega}
\end{equation}

The zero-point energy can now be minimized numerically, e.g.\ by using
simulated annealing.  However, it turns out that the states minimizing
Eq.~(\ref{eq-E1}) retain a large residual degeneracy.  Hence,
finding one such state does not yet solve the problem.  We must first
understand the nature of the residual degeneracy.  This is best done
by discussing a hidden symmetry responsible for it.

{\em A gauge-like symmetry.} Observe that the triangles $\{\alpha\}$
occupy the sites of a honeycomb lattice dual to the kagome
[Fig.~\ref{fig-wafer}(b)].  The in-plane sites $\{\alpha\beta\}$ live
on links of the dual lattice.  It was noted by Henley\cite{Henley}
that Eq.~(\ref{eq-omega}) reveals a special gauge-like symmetry
of the zero-point energy (\ref{eq-E1}).  Namely, a
transformation
\begin{equation}
\sigma_\alpha \mapsto \sigma_\alpha' = \Lambda_\alpha \sigma_\alpha,
\hskip 5mm
s_{\alpha\beta} \mapsto s_{\alpha\beta}' 
= \Lambda_\alpha s_{\alpha\beta} \Lambda_{\beta}^{-1}
\label{eq-gauge}
\end{equation}
leaves Eq.~(\ref{eq-omega}) invariant.  (Conservation of Ising spin
length $|s_{\alpha\beta}|$ requires that $\Lambda_\alpha = \pm 1$.)
Therefore, two spin configurations $\{s_{\alpha\beta}\}$ and
$\{s_{\alpha\beta}'\}$ related by the gauge symmetry (\ref{eq-gauge})
have identical magnon spectra and, consequently, equal zero-point
energies.  We therefore expect a large residual degeneracy of the
ground state even after the inclusion of the first quantum correction
(\ref{eq-E1}).

A word of caution.  Transformation (\ref{eq-gauge}) is not exactly a
gauge symmetry, even though it strongly resembles one.  The
transformed state must respect the ground-state constraint
(\ref{eq-constraint}).  The importance of this restriction is evident
when one ponders the meaning of the transformation: if
$\Lambda_\alpha=-1$, we flip {\em all} Ising spins on triangle
$\alpha$.  Doing so may violate the ground-state constraint
(\ref{eq-constraint}) on adjacent triangles, where only one spin is
flipped.  This reduces the number of possible gauge transformations
but---crucially---does not eliminate them entirely.

The uncovering of a hidden symmetry greatly simplifies our task: as
soon as we find a single ground state minimizing the zero-point energy
(\ref{eq-E1}), we can generate all of the others by applying the gauge
transformations (\ref{eq-gauge}).  But the gauge-like symmetry is
useful in more ways than one.  It also puts restrictions on the way in
which the zero-point energy depends on the Ising spin variables.

From the viewpoint of the $Z_2$ symmetry (\ref{eq-gauge}), the Ising
spins $s_{\alpha\beta}$ are gauge variables.  In a true gauge theory,
the energy can only depend on their gauge-invariant combinations, in
this instance the $Z_2$ fluxes through closed loops $\Gamma_\ell$ of
various lengths $\ell$ on the hexagonal lattice, $\phi_{\gamma_\ell} =
s_{\alpha_1\alpha_2} \, s_{\alpha_2\alpha_3} \ldots
s_{\alpha_\ell\alpha_1}$.  The flux through any such loop can be
factored into a product of fluxes piercing elementary plaquettes of
the dual lattice, the hexagons $\gamma = \Gamma_6$
[Fig.~\ref{fig-wafer}(b)].  Therefore, the zero-point energy will be a
function of $Z_2$ fluxes:
\begin{equation}
E_1 = E_1(\{\phi_{\gamma}\}).
\label{eq-E1-flux}
\end{equation}

While ours is not precisely a gauge theory, Eq.~(\ref{eq-E1-flux}) is
the right guess: {\em the zero-point energy is entirely determined by the
values of the $Z_2$ fluxes.}  To prove this, we demonstrate that any
two collinear ground states with the same set of flux variables
$\{\phi_{\gamma}\}$ have equal zero-point energies (\ref{eq-E1}).  To
accomplish that, we show that the two states are related by a
gauge-like transformation (\ref{eq-gauge}).

{\em Proof.}  Let the Ising link variables be $\{s_{\alpha\beta}\}$
and $\{s_{\alpha\beta}'\}$ in the two states.  Define new link
variables as their ratio, $t_{\alpha\beta} =
s_{\alpha\beta}'/s_{\alpha\beta} = \pm 1$.  The variables
$t_{\alpha\beta}$ are pure gauge: their product along any closed loop
equals $+1$.  Therefore we can unambiguously define a new variable
$\Lambda_\alpha$ on sites of the dual lattice in such a way that
$\Lambda_\alpha = t_{\alpha\beta} \Lambda_\beta$ everywhere.  Hence
$s_{\alpha\beta}' = \Lambda_\alpha s_{\alpha\beta}
\Lambda_\beta^{-1}$.  By Eq.~(\ref{eq-gauge}), the zero-point energies
of the two states are equal.  Q.E.D.

The precise form of the flux potential is not fixed by
Eq.~(\ref{eq-E1-flux}) and can be rather complicated.  One might hope
that a series expansion could successfully mimic it:
\begin{equation}
\frac{E_1}{N} = a_0 + \frac{1}{N} \sum_{\gamma}
a_1 \phi_\gamma 
+ \frac{1}{2 N^2} \sum_{\gamma,\gamma'}
a_2(\gamma, \gamma') \, \phi_\gamma \, \phi_{\gamma'} + \ldots 
\label{eq-cluster}
\end{equation}
If the linear term dominates, the ground state has uniform flux
$\phi_\gamma = -\mathrm{sign}(a_1)$
on all elementary plaquettes.  On the checkerboard, $\phi_\gamma = +1$
in the ground state.\cite{Tch03} 

Extending the same simplistic approach to the pyrochlore slice we can
take a guess that the ground states will either have no flux
($\phi_\gamma = +1$) or have the maximal flux $\pi$ ($\phi_\gamma = -1$).
While neither of these is the correct answer, they are interesting in
their own right.

{\em The $\pi$-flux states.}  In these ground states, the product of
six Ising spins around any hexagon (of the original kagome lattice)
equals $-1$ [Fig.~\ref{fig-wafer}(a)].  The $\pi$ flux on the hexagon
is related to the number of frustrated bonds.  (A bond is frustrated
if it connects Ising spins $s_i = s_j$.)  More precisely, each hexagon
has 0 or 2 frustrated bonds belonging to the triangles pointing {\em
down} [Fig.~\ref{fig-wafer}(b)].  This fact allows us to map any
$\pi$-flux state onto a dimer covering.  If a hexagon has 2 such
frustrated bonds, draw a dimer connecting the centers of the two
triangles contributing the frustrated bonds [Fig.~\ref{fig-wafer}(c)].
Because every triangle has exactly one frustrated bond, it shares a
dimer with one other triangle.  The same construction applies to the
sublattice of triangles pointing up [Fig.~\ref{fig-wafer}(d)]. We note
that gauge transformations on the down triangles are sufficient to
take any of the dimer states on the up triangles into any other
without changing the dimer state on the down triangles -- and vice
versa.  Thus every $\pi$-flux ground state maps onto a pair of {\em
completely independent} dimer coverings of the triangular lattices.

Now one can use the well-known properties of the dimer model on the
triangular lattice\cite{Fendley} to characterize the ensemble of
the $\pi$-flux states.  By construction, the number of such
states equals the number of pairs of dimer coverings.  The latter
grows exponentially with the number of sites, so that there is a
finite entropy of $0.4286$ per tetrahedron.  A disordered state of
classical dimers on a triangular lattice translates into the absence
of valence-bond order in our spin model.  Spin correlations are also
short-ranged.  On a lattice with periodic boundary conditions (on a
torus), these ground states fall into four disconnected topological
classes distinguished by the values of the global $Z_2$ fluxes
piercing the torus.

In short, the ensemble of $\pi$-flux states describes a classical
valence-bond liquid.

{\em The actual ground states.}  We have performed a numerical
minimization of the zero-point energy (\ref{eq-E1}) in the space of
collinear vacua (\ref{eq-constraint}) using the method of simulated
annealing.  A Monte Carlo step, conserving the classical energy
(\ref{eq-H}), consisted of flipping a randomly chosen chain of
alternating spins.  The zero-point energy (\ref{eq-E1}) was then 
determined numerically.

\begin{figure}
\centerline{\includegraphics[width=\columnwidth]{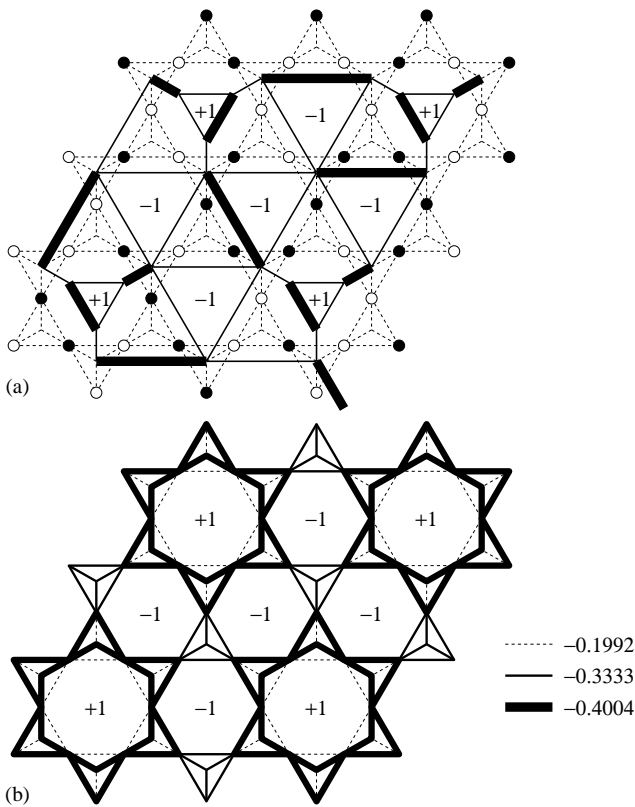}}
\caption{(a) A generic $2\times2$ collinear ground state and the
corresponding dimer covering (only one dual sublattice is shown).
Solid lines depict the decorated triangular lattice.  The numbers
$\pm1$ indicate the $Z_2$ flux.  (b) Expectation values of the
valence-bond variables $\langle s_i s_j\rangle$ are encoded in line
thickness.}
\label{fig-vbc}
\end{figure}

Initial runs were done on systems containing between 100 and 200
tetrahedra.  Starting from a periodic state we randomized the system
by flipping randomly chosen antiferromagnetic chains until spin
correlations with the original state were lost.  The lattice was then
annealed to achieve a state with the lowest energy.  Throughout a
simulation, we monitored Ising spins $s_i$, valence-bond variables
$s_i s_j$, and $Z_2$ fluxes on hexagons.  The low-energy states
obtained by the annealing showed no apparent spin or valence-bond
order [Fig.~\ref{fig-vbc}(a)].  However, the $Z_2$ fluxes showed a
clear tendency to order: 1/4 quarter of the hexagons had the trivial
$Z_2$ flux $+1$ and formed a regular pattern [Fig.~\ref{fig-vbc}(a)].
We denote such states as $2\times2$ after the size of the unit cell.

Finite-size scaling studies (with up to $10^4$ tetrahedra) yielded the
zero-point energies of $-1.2127 JS$, $-1.2462 JS$, and $-1.2497 JS$
per tetrahedron for the no-flux, $\pi$-flux, and $2\times2$ states,
respectively, with the $2\times2$ states the lowest.  On the basis of
the energy differences, we estimate that quantum selection of ground
states will be effective below the temperature scale of roughly
$10^{-2} JS/k_B$.  Note that the na\"\i ve energy scale $JS$ comes
with a numerically small prefactor.

To obtain a dimer model appropriate for the $2\times2$ flux pattern,
we start with the dimers on a triangular lattice, which yield the
$\pi$-flux states [Fig.~\ref{fig-wafer}(c)].  To change the flux from
$-1$ to $+1$ on one quarter of the hexagons, we decorate the
triangular lattice by modifying one quarter of triangles
[Fig.~\ref{fig-vbc}(a)].  Again, every frustrated bond is intersected
by a dimer.

The properties of classical dimers on the decorated lattice can be
readily computed by using Grassmann variables. The
number of $2\times2$ ground states is exponentially large, with
an entropy per tetrahedron of $0.4248$.  The probability to find a
dimer on a short link is $p = 0.4004 > 1/3$.  In spin language this
means that
$\langle s_i s_j
\rangle = p(+1) + (1-p)(-1) = -0.1992 > -1/3$ on links surrounding the
hexagons marked $+1$.  These bonds are therefore more frustrated than
the lattice average.  On the other hand, a tetrahedron placed
symmetrically with respect to the three closest fluxes $+1$ has 
$\langle s_i s_j \rangle = -1/3$ for all its bonds.  The
valence-bond averages $\langle s_i s_j \rangle$ are shown
schematically in Fig.~\ref{fig-vbc}(b).  They are nonuniform,
therefore the ensemble of $2\times2$ states has valence-bond order
that enlarges the lattice unit cell.  (We note
in passing a similarity with the valence-bond order proposed earlier
for the three-dimensional $S=1/2$
pyrochlore:\cite{Harris91,Tsunetsugu01etal} 
3/4 of tetrahedra have unequal
valence-bond correlations; the other 1/4 remain symmetric.)

Spin correlations $\langle s_i s_j \rangle$ between more distant
neighbors decay exponentially with distance.  The correlation length
is 1.15 (the unit of length is the side of a tetrahedron).  This
points to the absence of N\'eel order.

To summarize, we find that, to leading order in $1/S$, quantum
fluctuations select a collinear ensemble of degenerate ground states with
additional bond order and no spin order: we have a valence-bond
crystal.  This degeneracy may be lifted in higher orders in $1/S$,
as found previously for the Heisenberg antiferromagnet on 
kagome.\cite{Chubukov92,Henley95} 

The authors thank A.G. Abanov, C. Broholm, C.L. Henley, S.L. Sondhi,
and O.A. Starykh for useful discussions. R. M. was 
supported in part by the Minist\`ere de la Recherche et des Nouvelles
Technologies.

\end{document}